\documentclass[a4]{article}
\usepackage[dvips]{graphicx}

\begin{document}

\begin{titlepage}
\begin{flushright}
TU- 699\\
to appear in Phys. Rev. A69, 052114 (2004)\\
\end{flushright}
\ \\
\begin{center}
\LARGE
{\bf
Impossibility of distant indirect measurement of the quantum Zeno effect
}
\end{center}
\ \\
\begin{center}
\Large{
M.Hotta${}^\ast$ and M.Morikawa${}^\dagger$    }\\
{\it
${}^\ast$
Department of Physics, Faculty of Science, \\ 
Tohoku University, Sendai 980-8578,Japan\\
hotta@tuhep.phys.tohoku.ac.jp \\

\ \\

${}^\dagger$
Physics Department, Ochanomizu University,\\
Tokyo, 112-8610, Japan \\
hiro@phys.ocha.ac.jp
}
\end{center}

\begin{abstract}
We critically study the possibility of the quantum Zeno effect for indirect measurements.  
If the detector is prepared to detect the emitted signal from the core system, 
and does not reflect this signal back to the core system, 
then we can prove that the decay probability of the system is not changed by the continuous 
measurement of the signal and the quantum Zeno effect would never take place.   
This argument also applies to the quantum Zeno effect for accelerated two-level systems and 
unstable particle decays.  

\end{abstract}

\end{titlepage}

\section{Introduction}

\ \newline

Continuous measurement of a quantum system freezes the dynamics. \ This
quantum Zeno effect provides extreme nature of quantum mechanics\cite%
{misra77}\cite{itano90}. \ There are three kinds of causes of the quantum
Zeno effect; (1) continuous measurement of the system, (2) the effect of the
environment surrounding the system, and (3) the renormalization effect due
to the interaction between the system and the detectors. \ In any case, the
freezing of the total Hilbert space into several subspaces\cite{facchi03} is
the essence of the mechanism; a strong external disturbance dominates the
total Hamiltonian whose eigenspaces form the subspaces. \ Among the above
three causes, the last two can be described by (effective) Hamiltonian and
therefore are relatively well understood. However, the first one relies on
the phenomenological projection postulate a la von Neumann and needs a more
careful investigation. \ In particular, the consistency between the global
nature of the projection postulate and the causality, in the sense that any
signal cannot exceed the light velocity $c$, is problematic. \ This issue
will become particularly evident in the indirect quantum Zeno effect. \ 

For a simple example of the distant indirect measurement, suppose the
excited state of a two-level system of size $d$ decays to the ground state
by emitting light, which is continuously monitored by a detector located at $%
l(>d/2)$. \ (See Fig.1.)

\begin{figure}[h]
\begin{center}
\includegraphics[width=5cm,clip]{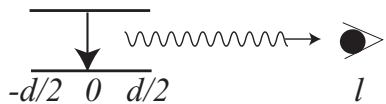}
\end{center}
\caption{Schematic description of the distant indirect measurement. The
excited state of a two-level system of size $d$ decays to the ground state
by emitting light, which is continuously monitored by the detector located
at $l(>d/2)$.}
\label{1}
\end{figure}
How and when does the quantum Zeno effect take place? Or does the effect
never take place? \ Considering both the projection postulate and the
causality, we may naturally expect that the Zeno effect, provided the decay
law is appropriate for it, takes place only after the time $(2l-d)/c$. \ If
this is the case, then what would happen when the two-level system is
constantly accelerated and the detectors are set outside the causal boundary
such that they can never affect the system? \ In Fig.2., we show this
situation schematically. \ The trajectory of the two-level system (the solid
line) is described by the hyperbolic function which approaches to the
``horizon'' $ct=x$ asymptotically in the future. The excited state of the
two-level system decays to the ground state by emitting light. In order to
monitor this signal, we prepare many detectors which are aligned so that
their trajectories are entirely located above this horizon. The causality
principle prevents such devices from affecting the two-level system through
physical interaction.(See Fig.2.) \ Then one could ask the questions we
posed at the beginning of the previous paragraph. The measurement process
may be able to make the system wave function change beyond causality because
the ``wave function'' itself has no physical reality. \ However since the
``probability'' is a physical reality, the causality may prohibit the Zeno
effect from taking place. \ Therefore, the problem here is the fact that, in
this case, the change in ``wave function'' describes that of
``probability'', and thus the possibility of the Zeno effect is very
problematic. \ 

\begin{figure}[h]
\begin{center}
\includegraphics[width=5cm,clip]{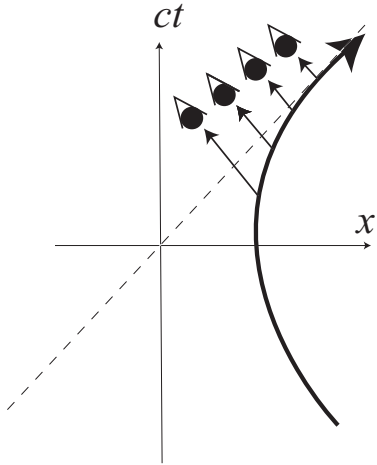}
\end{center}
\caption{Schematic description of the indirect measurement of an accelerated
two-level system in the space-time diagram. The trajectory of the two-level
system (the solid line) is described by the hyperbolic function which
approaches to the ``horizon'' $ct=x$ asymptotically in the future. All the
activated detectors are prepared so that their trajectories are entirely
located above this horizon. Causality principle prevents such devices from
affecting the system through physical interaction. }
\label{SNS2}
\end{figure}

This distant indirect measurement is very common in high-energy particle
experiments. \ Especially the issue becomes prominent in the decay
experiments of unstable particles. \ Can we expect that the particle decay
law is modified by the continuous monitoring of the emitted photon
associated with the decay, by detectors surrounding the unstable particle? \
Moreover the issue becomes much prominent in macroscopic situations. \ Is it
possible to expect that the life of the Schr\"{o}dinger's cat\footnote{%
The Schr\"{o}dinger's cat system is the strongly entangled state of the
microscopic unstable atom and the macroscopic life. \ } is elongated by the
continuous measurement of the system from outside? \ 

In this paper, we study this consistency issue between the global nature of
the von Neumann projection postulate in quantum mechanics and the causality
principle in relativity specifying the case of indirect measurements. \ 

This paper is organized as follows. \ In section two, we specify, in the
simplest form, the general feature of the indirect quantum Zeno experiments,
and prove, under some conditions, the impossibility of this effect in the
indirect distant measurement. \ Based on this argument, in section three, we
study a simple model which demonstrates the situation. \ Finally, in the
last section of this paper, we discuss several other applications of our
argument. \ 

\section{\protect\bigskip Distant Indirect Measurements}

\ Let us consider a quantum system with unitary evolution. The total system
may actually be very complicated however a subspace $\mathcal{H}_{Z}$ which
is relevant to consider the quantum Zeno effect would be almost closed in
dynamics. We would like to concentrate on this subspace $\mathcal{H}_{Z}$ .
Then one can always construct an experiment of quantum Zeno effect using
indirect measurement, as will be explained later in detail. \ 

Let $\{|e\rangle ,\{|n\rangle \}_{n=1,2,...}\}$ denote the complete
orthonormal basis vectors in $\mathcal{H}_{Z}$. In the following we
concentrate on the survival probability 
\begin{equation}
s(t)=|\langle e(0)|e(t)\rangle |^{2}
\end{equation}
of a state $|e\rangle $ in $\mathcal{H}_{Z}$, where $|e(t)\rangle $ is the
state evolved from $|e\rangle $ ($|e\rangle =$ $|e(0)\rangle $). Since the
evolution of the state vector is closed in this space, 
\begin{equation}
|e(t)\rangle =C(t)|e\rangle +\sum_{n}C_{n}(t)|n\rangle ,
\end{equation}
the survival probability $s(t)=|C(t)|^{2}$ can be obtained without direct
measurement of the state $|e\rangle $. Actually, the measurement of the
states $|n\rangle $ provides us with the probability $s(t)$ through the
unitary relation: 
\begin{equation}
s(t)=1-\sum_{n}|C_{n}(t)|^{2}.
\end{equation}
This is the indirect measurement of $s(t)$, which is the subject of our
investigation. Due to the time reversal symmetry of the Schr\"{o}dinger
equation, for short time intervals, $s(t)$ behaves like 
\begin{equation}
s(\Delta t)\sim 1-\alpha \Delta t^{2},
\end{equation}
where $\alpha $ is a constant. Therefore, at time $t=N\Delta t$ and after N
successive measurements the survival probability of the state $|e\rangle $
is calculated to be 
\begin{equation}
s_{N}(t)\sim \left( 1-\alpha (\Delta t)^{2}\right) ^{\frac{t}{\Delta t}}\sim
e^{-\alpha t\Delta t}.
\end{equation}
Consequently, for the continuous measurements of the states $|n\rangle $( in
the limit $N\rightarrow \infty $, $\Delta t\rightarrow 0$ with $t$ fixed)
one obtains 
\begin{equation}
\lim_{N\rightarrow \infty }s_{N}(t)=1,
\end{equation}
and thus, the time evolution of the state $|e\rangle $ freezes. This is the
essence of the quantum zeno effect with indirect measurement\footnote{%
Koshino and Shimizu\cite{koshino02} investigated a model of the indirect
measurements of the Zeno effects and found even an incomplete measurement is
sufficient to yield the effect. \ Our argument is different from theirs in
an essential point. \ See the end of discussion for more detail. \ 
\par
{}}.

For a much elaborate argument of the indirect measurement, let us suppose
the subspace $\mathcal{H}_{Z}$ can be decomposed into mutually complementary
subspaces $\mathcal{H}_{C}$ and $\mathcal{H}_{W}$ ($\mathcal{H}_{Z}=\mathcal{%
H}_{C}\oplus \mathcal{H}_{W}).$ \ We call $\mathcal{H}_{C}$ as the core-zone
subspace and $\mathcal{H}_{W}$ as the wave-zone subspace if they satisfy the
following property: \ 

(I) Any vector $|W\rangle $ in $\mathcal{H}_{W}$ remains in $\mathcal{H}_{W}$
as it evolves at any advanced time: 
\begin{equation}
|W(t>0)\rangle :=U_{+}(t)|W\rangle \in \mathcal{H}_{W},
\end{equation}
where $U_{+}(t>0)=e^{-\frac{i}{\hbar }tH}$ is the advanced time evolution
operator for the system and $H$ is the Hamiltonian of the system.

If we introduce the projection operators $P_{C}$, $P_{W}$, and the unit
projection operator $\mathbf{1}_{Z}$ which project a state of $\mathcal{H}%
_{Z}$ onto $\mathcal{H}_{C}$, $\mathcal{H}_{W},$ and $\mathcal{H}_{Z}$,
respectively, where 
\begin{equation}
P_{C}+P_{W}=\mathbf{1}_{Z},
\end{equation}
the property (I) can also be expressed as the followings:

(II) For any wave-zone vector $|W\rangle $, 
\begin{equation}
P_{C}|W(t>0)\rangle =P_{C}U_{+}(t)|W\rangle =0,
\end{equation}
or,

(III)For $t>0$, 
\begin{equation}
P_{C}U_{+}(t)P_{W}U_{+}(t)^{\dagger }=0.  \label{13}
\end{equation}

Clearly, the emergence of the subspace $\mathcal{H}_{W}$ with a nontrivial $%
\mathcal{H}_{C}$ is allowed only in the case when $\mathcal{H}_{W}$ is an
open system. In this paper we call $|C\rangle (\in \mathcal{H}_{C})$ a
core-zone state and $|Z\rangle (\in \mathcal{H}_{W})$ a wave-zone state. \
From any of the properties (I), (II), or (III), we obtain the following
useful lemma. \ 

Lemma: If two state vectors $|\Psi _{1}(t_{o})\rangle $ and $|\Psi
_{2}(t_{o})\rangle $ in $\mathcal{H}_{Z}$ satisfy the following relation at
time $t_{o}$, 
\begin{equation}
P_{C}|\Psi _{1}(t_{o})\rangle =P_{C}|\Psi _{2}(t_{o})\rangle ,
\label{lemma1}
\end{equation}
then this relation always holds at time $t(>t_{o})$ in the future: 
\begin{equation}
P_{C}|\Psi _{1}(t)\rangle =P_{C}|\Psi _{2}(t)\rangle .
\end{equation}

To prove this statement, let us define a core-zone state $|C\rangle $ as 
\[
|C\rangle :=P_{C}|\Psi _{1}(t_{o})\rangle =P_{C}|\Psi _{2}(t_{o})\rangle . 
\]
Then using two wave-zone states $|W_{1}\rangle $ and $|W_{2}\rangle $, the
initial states can be expressed as 
\begin{equation}
|\Psi _{1}(t_{o})\rangle =|C\rangle +|W_{1}\rangle ,
\end{equation}
\begin{equation}
|\Psi _{2}(t_{o})\rangle =|C\rangle +|W_{2}\rangle .
\end{equation}
Next, let us evolve the two states until the time $t$. 
\begin{equation}
|\Psi _{1}(t)\rangle =U_{+}(t-t_{o})|C\rangle +U_{+}(t-t_{o})|W_{1}\rangle ,
\label{1}
\end{equation}
\begin{equation}
|\Psi _{2}(t)\rangle =U_{+}(t-t_{o})|C\rangle +U_{+}(t-t_{o})|W_{2}\rangle .
\label{2}
\end{equation}
Because $|W_{1}\rangle $ and $|W_{2}\rangle $ are wave-zone states and $%
t-t_{o}>0$ holds, they satisfy 
\begin{equation}
P_{C}U_{+}(t-t_{o})|W_{1}\rangle =P_{C}U_{+}(t-t_{o})|W_{2}\rangle =0.
\end{equation}
according to the property (II). \ Thus by operating $P_{C}$ in eqns. (\ref{1}%
) and (\ref{2}), we obtain the result mentioned in the lemma: 
\begin{equation}
P_{C}|\Psi _{1}(t)\rangle =P_{C}U_{+}(t-t_{o})|C\rangle =P_{C}|\Psi
_{2}(t)\rangle .
\end{equation}

Now let us take a core-zone state $|e\rangle $ as the initial state at $t=0$%
. We calculate $s_{N}(t)$, which is the survival probability of $|e\rangle $
at a later time $t$ after N successive $\mathcal{H}_{W}$ measurements and
compare it with the same quantity without any $\mathcal{H}_{W}$ measurements 
$s(t)(=|\langle e(0)|e(t)\rangle |^{2})$. Here, we assume that all the
measurements are completely ideal and represented by the von Neumann
projection postulate. Let us explain the procedure in more detail. The
initial state $|e(0)\rangle $ is set at time $t=0$. The measurements on $%
\mathcal{H}_{W}$ are performed at times $t=t_{i}$ ($i=1\sim N$ and $%
0<t_{1}<t_{2}<\cdots <t_{N}$). Each measurement makes the wave function
shrink in either of the two ways, depending upon whether the observed state
of the system belongs to $\mathcal{H}_{W}$ or not. After the N-times
measurements, we finally check at time $t=t_{N+1}(>t_{N})$ whether the
decaying two-level atom is still in the initial excited state $|e(0)\rangle $
or not. Repeating the procedure many times would yield the observational
value $_{{}}$of the survival probability $s_{N}(t)$.

From $t=0$ to $t=t_{1}$ the state vector evolves by the unitary evolution $%
U_{+}(t)$ as 
\begin{equation}
|\Psi (t)\rangle =U_{+}(t)|e(0)\rangle =|e(t)\rangle .
\end{equation}
At time $t=t_{1}$, the first measurement is executed. The probability of
finding that the observed state belongs to $\mathcal{H}_{W}$ is given as 
\begin{equation}
p_{1}=\langle \Psi (t_{1})|P_{W}|\Psi (t_{1})\rangle .
\end{equation}
If this case is realized, the state shrinks into a wave-zone state $|\Psi
_{\lbrack w]}(t_{1})\rangle $ defined by 
\begin{equation}
|\Psi _{\lbrack w]}(t_{1})\rangle =\frac{1}{\sqrt{p_{1}}}P_{W}|\Psi
(t_{1})\rangle .
\end{equation}
Because $|\Psi _{\lbrack w]}(t_{1})\rangle \in \mathcal{H}_{W}$ and any
wave-zone state evolves only within $\mathcal{H}_{W}$ (I), the system is no
longer found in any state of $\mathcal{H}_{C}$ in the remaining
measurements. Hence $|\Psi _{\lbrack w]}\rangle $ does not contribute at all
to the survival probability of $|e(0)\rangle $, which is a core-zone state,
and we can completely ignore this possibility in the calculation of $%
s_{N}(t) $. If one does not find any state in $\mathcal{H}_{W}$, the state
shrinks into a core-zone state $|\Psi _{\lbrack c]}(t_{1})\rangle $ defined
by 
\begin{equation}
|\Psi _{\lbrack c]}(t_{1})\rangle =\frac{1}{\sqrt{1-p_{1}}}P_{C}|\Psi
(t_{1})\rangle .  \label{9}
\end{equation}
The probability of this case is given by $1-p_{1}$. Here the projective
relation $P_{C}^{2}=P_{C}$ directly yields the following equation: 
\begin{equation}
P_{C}|\Psi _{\lbrack c]}(t_{1})\rangle =\frac{1}{\sqrt{1-p_{1}}}P_{C}|\Psi
(t_{1})\rangle .
\end{equation}
Due to the above lemma, we obtain the following relation at the future time $%
t=t_{2}(>t_{1})$: 
\begin{equation}
P_{C}|\Psi _{\lbrack c]}(t_{2})\rangle =\frac{1}{\sqrt{1-p_{1}}}P_{C}|\Psi
(t_{2})\rangle .
\end{equation}
Consequently, it is possible to write the probability of finding that the
state of the system belongs to $\mathcal{H}_{C}$ at $t=t_{2}$ if it belonged
to $\mathcal{H}_{C}$ at $t=t_{1}$ as follows. 
\begin{equation}
1-p_{2}=1-\langle \Psi _{\lbrack c]}(t_{2})|P_{W}|\Psi _{\lbrack
c]}(t_{2})\rangle .
\end{equation}
Thus the probability of not finding the system in any state of $\mathcal{H}%
_{W}$ in both measurements at $t=t_{1}$ and $t=t_{2}$ is given by 
\begin{equation}
(1-p_{1})\times (1-p_{2}).
\end{equation}
If this situation is realized, the state shrinks into the core-zone state
defined by 
\begin{equation}
|\Psi _{\lbrack cc]}(t_{2})\rangle =\frac{1}{\sqrt{1-p_{2}}}P_{C}|\Psi
_{\lbrack c]}(t_{2})\rangle .  \label{3}
\end{equation}
By virtue of the above lemma it is easily shown that 
\begin{equation}
P_{C}|\Psi _{\lbrack cc]}(t_{3})\rangle =\frac{1}{\sqrt{1-p_{2}}}P_{C}|\Psi
_{\lbrack c]}(t_{3})\rangle .  \label{4}
\end{equation}
Because the relation $t_{3}>t_{1}$ holds, one can also write 
\begin{equation}
P_{C}|\Psi _{\lbrack c]}(t_{3})\rangle =\frac{1}{\sqrt{1-p_{1}}}P_{C}|\Psi
(t_{3})\rangle .  \label{5}
\end{equation}

We can repeat the same procedure $N$-times. \ For each measurement
procedure, in the similar way in eqns. (\ref{9}) and (\ref{3}), we define
the shrunk states as 
\begin{equation}
|\Psi _{\lbrack c^{\otimes n}]}(t_{n})\rangle =\frac{1}{\sqrt{1-p_{n}}}%
P_{C}|\Psi _{\lbrack c^{\otimes (n-1)}]}(t_{n})\rangle ,
\end{equation}
where $n=1\sim N$ and $p_{n}$ is defined by 
\[
p_{n}=\langle \Psi _{\lbrack c^{\otimes (n-1)}]}|P_{W}|\Psi _{\lbrack
c^{\otimes (n-1)}]}\rangle . 
\]
Moreover note that by using the above lemma, we can prove, similarly to
eqns. (\ref{4}) and (\ref{5}), that 
\begin{equation}
P_{C}|\Psi _{\lbrack c^{\otimes k}]}(t_{N+1})\rangle =\frac{1}{\sqrt{1-p_{k}}%
}P_{C}|\Psi _{\lbrack c^{\otimes (k-1)}]}(t_{N+1})\rangle  \label{6}
\end{equation}
for $k=1\sim N$. As the last step at time $t=t_{N+1}$, we probe directly the 
$\mathcal{H}_{C}$ sector and determine the probability of finding the system
in the initial state, $s_{N}(t)$. The explicit form of $s_{N}(t)$ is given
by 
\[
s_{N}(t=t_{N+1})=\left[ \prod_{k=1}^{N}(1-p_{k})\right] \times \left|
\langle e(0)|\Psi _{\lbrack c^{\otimes N}]}(t_{N+1})\rangle \right| ^{2}. 
\]
The first factor in the right-hand-side is the probability of finding the
state is the core-zone state $|\Psi _{\lbrack c^{\otimes N}]}\rangle $ after
the $N$ measurements. The second factor is the transition probability from $%
|\Psi _{\lbrack c^{\otimes N}]}\rangle $ to $|e(0)\rangle $. It should be
reminded here that $|e(0)\rangle $ is a core-zone state, that is, 
\begin{equation}
\langle e(0)|=\langle e(0)|P_{C}.  \label{7}
\end{equation}
Using eqns.(\ref{6}) and (\ref{7}), we can rewrite $s_{N}$ as 
\begin{eqnarray}
s_{N}(t=t_{N+1}) &=&\prod_{k=1}^{N}(1-p_{k})\times \left| \langle
e(0)|P_{C}|\Psi _{\lbrack c^{\otimes N}]}(t_{N+1})\rangle \right| ^{2} 
\nonumber \\
&=&\prod_{k=1}^{N-1}(1-p_{k})\times \left| \langle e(0)|P_{C}|\Psi _{\lbrack
c^{\otimes (N-1)}]}(t_{N+1})\rangle \right| ^{2}  \nonumber \\
&\vdots &  \nonumber \\
&=&(1-p_{1})\times \left| \langle e(0)|P_{C}|\Psi _{\lbrack
c]}(t_{N+1})\rangle \right| ^{2}  \nonumber \\
&=&\left| \langle e(0)|\Psi (t_{N+1})\rangle \right| ^{2}=\left| \langle
e(0)|e(t_{N+1})\rangle \right| ^{2},
\end{eqnarray}
and therefore, we obtain the following theorem:\newline

Theorem: If we take a core-zone state $|e(0)\rangle $ as the initial state
at $t=0$, under any of the conditions (I), (II) or (III), N-times
measurements on the wave-zone subspace $\mathcal{H}_{W}$ does not affect the
survival probability of $|e(0)\rangle $ at all, i.e.: 
\begin{equation}
s_{N}(t)=s(t).  \label{theorem}
\end{equation}

It should be emphasized that we have not specified the time dependence of $%
s(t)$. The theorem is applicable even when the form of $s(t)$ is different
from the exponential form $\propto e^{-\Gamma t}$. Hence, even if s(t) obeys
the standard early behavior like $s(t)\sim 1-\alpha t^{2}$, the wave-zone
measurements do not yield the Zeno effect at all.

The above argument leading to the theorem can be summarized as follows. The
statement (I) means ``wave-zone states evolve within $\mathcal{H}_{W}$ in
the future'':$[P_{W},U_{+}]P_{W}=0.$ Since $P_{C}+P_{W}=\mathbf{1}_{Z},$
this leads to $[P_{C},U_{+}]P_{W}=0,$ which is equivalent to $%
P_{C}U_{+}(t)P_{W}=0.$ (These are other forms of (II)(III). )\ The last
relation yields the above lemma $P_{C}[P_{C},U_{+}]=0$, which claims ``Any
part of a state which finally evolves to a state in $\mathcal{H}_{C}$ has
been evolving within $\mathcal{H}_{C}$ entirely''. \ The repeated use of the
lemma in this form yields 
\begin{equation}
P_{C}U_{+}(\Delta t_{n})P_{C}U_{+}(\Delta t_{n-1})\cdots P_{C}U_{+}(\Delta
t_{1})=P_{C}\prod_{j=1}^{n}U_{+}(\Delta t_{j}),  \label{[2]}
\end{equation}
which clearly shows this statement. \ 

By inserting the normalization factor after each projection in eqn.(\ref{[2]}%
), 
\begin{eqnarray}
&&\frac{1}{\sqrt{1-p_{n}}}P_{C}U_{+}(\Delta t_{n})\frac{1}{\sqrt{1-p_{n-1}}}%
P_{C}U_{+}(\Delta t_{n-1})\cdots \frac{1}{\sqrt{1-p_{1}}}P_{C}U_{+}(\Delta
t_{1})  \nonumber \\
&=&P_{C}\prod_{j=1}^{n}\frac{1}{\sqrt{1-p_{j}}}U_{+}(\Delta t_{j}),
\label{repmes}
\end{eqnarray}
one obtains the resultant state after successive negative-result measurement
on $\mathcal{H}_{W},$ where $1-p_{j}$ is the probability that the state is
not in $\mathcal{H}_{W}$ at the time $t_{j}$. \ When we calculate the
survival probability $s_{N}(t),$ the above normalization factor totally
cancels with these probabilities. Thus the final survival probability \ is
not affected by the repeated measurements; this is the theorem expressed
above in eqn.(\ref{theorem}). This cancellation of the factors $\sqrt{1-p_{j}%
}$ simply reflects the fact that only a single intermediate state
contributes in the calculation of the survival probability. \ Since there is
no interference between different intermediate states in this case, it is
clear that the measurement cannot affect the result. \ This claim applies
not only to the survival probability with the initial condition $%
|e(0)\rangle $, but also any process which finally results in a state in $%
\mathcal{H}_{C},$ according to eqn.(\ref{[2]}).

In the above arguments for the Hamiltonian system, the boundary/initial
condition, which reflects the wave-zone property (I), has been essential. \
This one-sided property would actually be associated with the most distant
indirect measurements of quantum processes. \ In the next section, we study
a typical model of indirect measurement which falls into this category. \ 

\section{ A Model}

\ We have developed the general formalism which leads to the theorem of eqn.(%
\ref{theorem}) in the previous section. \ Now we would like to study a
simple example which possesses a wave-zone subspace and the theorem is
applicable rigorously. Let us suppose a one-dimensional space in which $x$
denotes its spatial coordinate, and set a two-level atom system of size $d$
in the region $[-d/2,d/2]$. To express the upper and lower energy level
states, we introduce a fermionic pair of annihilation and creation operators 
$a$ and $a^{\dagger }$: 
\begin{equation}
\{a,\ a^{\dagger }\}=1,
\end{equation}
\begin{equation}
\{a^{\dagger },\ a^{\dagger }\}=\{a,\ a\}=0.
\end{equation}
We also introduce a massless spinor field 
\begin{equation}
\Phi (x)=(\Phi _{R}(x),\Phi _{L}(x))^{T}
\end{equation}
and quantize it in the fermionic way: 
\begin{equation}
\{\Phi _{h}(x),\ \Phi _{h^{\prime }}^{\dagger }(x^{\prime })\}=\delta
_{hh^{\prime }}\delta (x-x^{\prime }),
\end{equation}
\begin{equation}
\{\Phi _{h}^{\dagger }(x),\ \Phi _{h^{\prime }}^{\dagger }(x^{\prime
})\}=\{\Phi _{h}(x),\ \Phi _{h^{\prime }}(x^{\prime })\}=0,
\end{equation}
where $h(=R,L)$ is the helicity of the field excitations. As in the
Coleman-Hepp model \cite{ch}, using the annihilation operator, the vacuum
state $|vac\rangle $ can be introduced as 
\begin{equation}
a|vac\rangle =0,
\end{equation}
\begin{equation}
\Phi _{h}(x)|vac\rangle =0.
\end{equation}
Then the excited state of the two-level atom is defined by 
\begin{equation}
|e\rangle =a^{\dagger }|vac\rangle .
\end{equation}
For the spinor field, we concentrate on the two particle states in which
only one R-helicity and one L-helicity particles exist. The state in which a
R-helicity particle stays at the position $x=x_{R}$ and a L-helicity
particle at $x=x_{L}$ is denoted by 
\begin{equation}
|x_{R},x_{L}\rangle =\Phi _{R}^{\dagger }(x_{R})\Phi _{L}^{\dagger
}(x_{L})|vac\rangle .
\end{equation}
Now let us write the Hamiltonian of the total system, which is composed of
three terms: 
\begin{equation}
H=H_{atom}+H_{\Phi }+H_{int}.  \label{totH}
\end{equation}
The first term $H_{atom}$ is the Hamiltonian of the free motion of the
two-level atom and is given by 
\begin{equation}
H_{atom}=\hbar \omega a^{\dagger }a,
\end{equation}
where the energy of the excited state is set to be $\hbar \omega $. The
second term $H_{\Phi }$ is the free Hamiltonian of the massless spinor field
and is defined by 
\begin{eqnarray}
H_{\Phi } &=&-i\hbar c\int_{-\infty }^{\infty }\Phi ^{\dagger }\sigma
_{3}\partial _{x}\Phi dx  \nonumber \\
&=&-i\hbar c\int_{-\infty }^{\infty }\left[ \Phi _{R}^{\dagger }\partial
_{x}\Phi _{R}-\Phi _{L}^{\dagger }\partial _{x}\Phi _{L}\right] dx,
\end{eqnarray}
where $\sigma _{3}$ is the third component of the Pauli matrix. If no
interaction term is added, the field Hamiltonian yields right-moving
particles for $h=R$ and left-moving particles for $h=L$ with the light
velocity. The third term $H_{int}$ in eqn.(\ref{totH}) expresses the
interaction between the two-level atom and the spinor field and is given by 
\begin{eqnarray}
H_{int} &=&\hbar \int_{-d/2}^{d/2}dx\int_{-d/2}^{d/2}dx^{\prime }  \nonumber
\\
&&\times \left[ g(x,x^{\prime })\Phi _{R}^{\dagger }(x)\Phi _{L}^{\dagger
}(x^{\prime })a+g(x,x^{\prime })^{\ast }a^{\dagger }\Phi _{L}(x^{\prime
})\Phi _{R}(x)\right] .  \label{16}
\end{eqnarray}
The interaction induced by the coupling $g(x,x^{\prime })$ is supposed to
take place only in the restricted spatial region defined by $x,x^{\prime
}\in (-d/2,d/2)$ and the excited state of the atom decays into two particle
states with different helicities. Note that even after adding the
interaction term in eqn.(\ref{16}), $|vac\rangle $ is still stable.

In this model, the subspace, whose complete basis is given by $\{|e\rangle
,|x_{R},x_{L}\rangle \}$ with $x_{R}\geq -d/2$ and $x_{L}\leq d/2$, can be
identified as a subspace $\mathcal{H}_{Z}$, because the evolution in this
space is closed: 
\begin{equation}
|\Psi (t)\rangle =C(t)|e\rangle +\int_{-d/2}^{\infty }dx_{R}\int_{-\infty
}^{d/2}dx_{L}F(x_{R},x_{L};t)|x_{R},x_{L}\rangle  \label{14}
\end{equation}
for an arbitrary vector $|\Psi \rangle (\in \mathcal{H}_{Z})$. From the Schr%
\"{o}dinger equation, the amplitudes obey the following equations: 
\begin{eqnarray}
&&i\partial _{t}C=\omega
C(t)+\int_{-d/2}^{d/2}dx_{R}\int_{-d/2}^{d/2}dx_{L}g(x_{R},x_{L})^{\ast
}F(x_{R},x_{L};t),  \label{1001} \\
&&(\partial _{t}+c\partial _{x_{R}}-c\partial
_{x_{L}})F(x_{R},x_{L};t)=-ig(x_{R},x_{L})C(t).  \label{1002}
\end{eqnarray}
It is possible to integrate eqn.(\ref{1002}) formally and the result can be
expressed as 
\begin{eqnarray}
F(x_{R},x_{L};t) &=&F_{o}(x_{R}-ct,x_{L}+ct)  \nonumber \\
&&-i\int_{0}^{t}g(x_{R}-ct+c\tau ,x_{L}+ct-c\tau )C(\tau )d\tau ,
\label{1003}
\end{eqnarray}
where $F_{o}(x_{R},x_{L})$ is the initial amplitude of $F$ at $t=0$.

When $x_{R}>d/2$ or $x_{L}<-d/2$, by taking the initial condition as 
\begin{eqnarray}
&&C(0)=0,  \label{54} \\
&&F_{o}(x_{R},x_{L})=\delta (x_{R}-x_{R}^{\prime })\delta
(x_{L}-x_{L}^{\prime }),  \label{55}
\end{eqnarray}
the following relation arises from eqn.(\ref{1001}) and eqn.(\ref{1003}); 
\begin{equation}
U_{+}(t)|x_{R},x_{L}\rangle =|x_{R}+ct,x_{L}-ct\rangle .  \label{56}
\end{equation}
This means that the right- and left-moving particles propagate freely after
leaving the interaction region. It is worth stressing that even if only one
of the two conditions $x_{R}>d/2$ or $x_{L}<-d/2$ holds, the evolution in
eqn.(\ref{56}) is still realized. This is because the interaction is
activated only when both particles simultaneously stay in the region $%
(-d/2,d/2)$.

From the above result, we can introduce in $\mathcal{H}_{Z}$ a wave-zone
subspace $\mathcal{H}_{W}$ for the excited state $|e\rangle $, which is
defined using the projection operator onto the subspace $P_{W}$: 
\begin{equation}
P_{W}=P_{+}+P_{-}+P_{+-},
\end{equation}

\begin{eqnarray}
P_{+} &=&\int_{d/2}^{\infty
}dx_{R}\int_{-d/2}^{d/2}dx_{L}|x_{R},x_{L}\rangle \langle x_{R},x_{L}|, \\
P_{-} &=&\int_{-d/2}^{d/2}dx_{R}\int_{-\infty
}^{-d/2}dx_{L}|x_{R},x_{L}\rangle \langle x_{R},x_{L}|, \\
P_{+-} &=&\int_{d/2}^{\infty }dx_{R}\int_{-\infty
}^{-d/2}dx_{L}|x_{R},x_{L}\rangle \langle x_{R},x_{L}|.
\end{eqnarray}
Then the core-zone subspace $\mathcal{H}_{C}$ is defined using the
projection operator: 
\[
P_{C}:=\mathbf{1}_{Z}-P_{W}=|e\rangle \langle
e|+\int_{-d/2}^{d/2}dx_{R}\int_{-d/2}^{d/2}dx_{L}|x_{R},x_{L}\rangle \langle
x_{R},x_{L}|. 
\]
By construction, we have 
\begin{equation}
P_{C}|e\rangle =|e\rangle .
\end{equation}
Now, the states are complete in $\mathcal{H}_{Z}$ as seen in eqn.(\ref{14}),
and the property (I) holds due to eqn.(\ref{56}), and therefore, everything
is in place to apply the theorem of the previous section. \ According to
this theorem, one can claim that the measurements of the $\Phi $ particles
in the outside regions $(-\infty ,-d/2]\cup \lbrack d/2,\infty )$ do not
affect the survival probability of the excited state of the atom at all,
hence, neither Zeno nor anti-Zeno effects take place.

On the other hand, if one observes the $\Phi $ particles not only in the
outside region but also in the inside interaction region $(-d/2,d/2)$, then
the development of $s_{N}(t)$ is expected to be modified by such
measurements. Clearly, when the continuous measurements of the $\Phi $
particles are performed in the whole region $(-\infty ,\infty )$, the
quantum Zeno effect will most strongly take place; $s_{N}\rightarrow 1$.

\section{Discussions and Comments}

The starting point of our argument has been the question of how the
causality and the projection postulate in quantum mechanics can be
reconciled with each other in the indirect quantum Zeno effects. \
Recognizing that the indirect distant measurement often possesses the
wave-zone property (I), we have studied this one-sided property, which led
to the lemma of eqn.(\ref{lemma1}). \ By calculating the survival
probability of a system under indirect measurement, we have found the
probability is not affected by the measurement at all. \ Further by using a
simple model, we have demonstrated the applicability of the general
argument. \ 

Some additional comments on our investigation are in order. \ \ \ 

The first one is related to the reflection wave of the detector. One may
suppose models in which the apparatus creates reflectional waves of the $%
\Phi $ field in the process of measurement. The reflected $\Phi $ wave
returns to the core-zone region and begins to affect the evolution of the $%
\mathcal{H}_{C}$ sector after the arrival time. Thus, the detector
Hamiltonian comes to violate the structure of the original wave-zone
property (I); the $\Phi $ particle states outside the atom are no longer
exact wave-zone states. Therefore our theorem is not exactly applicable in
such cases. However, the contribution of the reflectional wave would be
subleading in the coupling expansion and can be negligible in small coupling
cases as commented below again.\ 

The second comment is related to the completeness of the measurement. In the
above arguments, we have treated the ideal measurement just for convenience
of discussion. However our argument can be applicable for more realistic
incomplete measurements. \ For example, as in the work of Koshino and
Shimizu \cite{koshino02}, we can set the measurement apparatus outside the
atom ($x\in \lbrack x_{-},x_{+}]$), which measures the $\Phi $ field in the
above model with the following interaction: 
\begin{eqnarray}
H_{detector} &=&\hbar \int_{-\infty }^{\infty }\Omega (k)b^{\dagger
}(k)b(k)dk  \nonumber \\
&&+\hbar \int_{x_{-}}^{x_{+}}dx\int_{-\infty }^{\infty }dk\lambda
_{R}(x,k)\left( b^{\dagger }(k)\Phi _{R}(x)+\Phi _{R}^{\dagger
}(x)b(k)\right)  \nonumber \\
&&+\hbar \int_{x_{-}}^{x_{+}}dx\int_{-\infty }^{\infty }dk\lambda
_{L}(x,k)\left( b^{\dagger }(k)\Phi _{L}(x)+\Phi _{L}^{\dagger
}(x)b(k)\right) ,  \label{20}
\end{eqnarray}
where $x_{+}>x_{-}>d/2$, and $b(k)$ ($b^{\dagger }(k)$) is the annihilation
(creation) operator of the detector excitations: 
\[
\lbrack b(k),\ b^{\dagger }(k^{\prime })]=\delta (k-k^{\prime }), 
\]
and $\lambda _{h}(x,k)$ is a real coupling function which controls the
incompleteness of the detector. The third term of the Hamiltonian in eqn.(%
\ref{20}) generates a reflectional left-moving wave and the system may be in
two-left-moving-particle states. The time evolution in this case can be
expressed in a closed form as 
\begin{eqnarray}
|\Psi (t)\rangle &=&C(t)|e\rangle +\int_{-d/2}^{\infty }dx_{R}\int_{-\infty
}^{d/2}dx_{L}F(x_{R},x_{L};t)|x_{R},x_{L}\rangle  \nonumber \\
&&+\int_{-\infty }^{d/2}dx_{L}^{1}\int_{-\infty
}^{x_{+}}dx_{L}^{2}G(x_{L}^{1},x_{L}^{2};t)|x_{L}^{1},x_{L}^{2}\rangle 
\nonumber \\
&&+\int_{-\infty }^{d/2}dx_{L}\int_{-\infty }^{\infty
}dkD_{k}(x_{L};t)|b_{k}^{\dagger },x_{L}\rangle
\end{eqnarray}
where 
\[
|x_{L}^{1},x_{L}^{2}\rangle =\Phi _{L}^{\dagger }(x^{1})\Phi _{L}^{\dagger
}(x^{2})|vac\rangle , 
\]
and 
\[
|b_{k}^{\dagger },x_{L}\rangle =b_{k}^{\dagger }\Phi _{L}^{\dagger
}(x_{L})|vac\rangle . 
\]
Note that the interaction between the atom and the two left-moving particles
does not exist in the model. Therefore, the following relation is always
satisfied provided $x_{L}^{1},x_{L}^{2}<x_{-}$: 
\[
U_{+}(t)|x_{L}^{1},x_{L}^{2}\rangle =|x_{L}^{1}-ct,x_{L}^{2}-ct\rangle . 
\]
A crucial point is that the evolution of $C(t)$ and $F(x_{R},x_{L})$ for $%
x_{R},x_{L}\in (-d/2,d/2)$ in this case is exactly the same as in eqn.(\ref%
{1001}) and eqn.(\ref{1002}). Consequently, both the couplings $\lambda _{R}$
and $\lambda _{L}$ give no contribution to the survival probability of the
atom's excited state $s$ defined by 
\begin{equation}
s(t,\lambda _{h}]=\left| \langle e|\exp \left[ -\frac{i}{\hbar }t\left(
H+H_{detector}\right) \right] |e\rangle \right| ^{2}:
\end{equation}
\begin{equation}
\frac{\delta s(t,\lambda _{h}(x,k)]}{\delta \lambda _{h}(x,k)}=0.
\end{equation}
This implies that no quantum Zeno effect is observed even in this incomplete
measurement model with a reflectional wave.

Our argument will also be applicable for the decay experiments of unstable
particles with long life times. \ In the early study of the Zeno effect, it
has been argued that the unstable particle cannot decay when it is monitored
continuously. \ In general in this experiment, the light (or any signal)
emitted from the decayed particle is detected by the high sensitive
measuring device at distance, which is a typical case of the indirect
measurement. In the past analyses \cite{chiu82}, the focus has been on the
early time behavior of the survival probability $s(t)$. Especially it is
pointed out that the Zeno effect analysis inevitably requires the
nonperturbative behavior of $s(t)$ in the small-Q S-wave decay which is
still controversial due to the nonperturbative difficulty. However,
independent from their analysis, our theorem does not need any explicit time
dependence of $s(t)$ and can be quite useful. In the real three-dimensional
situation, most of the reflectional wave is expanded in space and only a
tiny part of it can go back to the decaying area. \ Further, this tiny part
has to reproduce the inverse-decay process followed by another decay process
in order to make the decay probability change. \ This process is at higher
order in small coupling constant and therefore makes the effect negligibly
small. \ Hence the property (I) would hold with fairly high precision. As a
result, this type of continuous measurement yields only negligible Zeno
suppression of the unstable particle decay, irrespective of the decay law's
detail. \ 

For the quantum Zeno effect in the constantly accelerated system, the
confrontation between the causality and the projection postulate has become
prominent, as explained in the introduction. \ According to our argument
however, the situation has become very clear and the paradox is settled. \
Since this case is the typical indirect measurement with the property (I)
which is guaranteed by the one-sided property of the causality horizon, the
measurement does not affect the decay probability at all. The same argument
would also apply to the evaporation of Black Holes due to the Hawking
radiation. \ \ 

When the size of the total system $\mathcal{H}_{Z}$ becomes macroscopic, the
property (I) tends to hold more naturally. \ For example in the \
macroscopic Scr\"{o}dinger's cat system, the external measurement of the cat
state would permit the property (I), and the quantum Zeno effect does not
take place; the continuous measurement does not change the destiny of the
cat. \ 

In the context of indirect measurement of the quantum Zeno effects, the work
by Koshino and Shimizu\cite{koshino02} is important. \ By investigating a
model of indirect measurements of Zeno and anti-Zeno effects in spontaneous
decay of a two-level atom, they found that the quantum Zeno effect really
takes place even if the detector of a photon excitation emitted from the
atom does not cover the full solid angle. \ We cannot however apply our
argument to their model because their detector is not spatially localized
and spreads over the space; the emitted photon continues to propagate in the
atom region until it is observed and therefore our property (I) does not
hold. \ On the other hand in our paper, we explicitly utilized the local
property of the (anti-)Zeno effect which was not considered in \cite%
{koshino02}. Actually the model we have used describes the case in which the
emitted excitations get out of the spatial region of the decaying two-level
atom at a finite time and propagate freely before the detection. \
Restricted in such a situation we have proved that the survival probability
of the decaying state is not affected at all by the successive indirect
measurements. \ 

\bigskip

\textbf{Acknowledgment}\newline

We would like to thank Akira Shimizu and Kazuki Koshino for useful
discussions and Mohammad Ahmady for correcting English of this manuscript. \

\end{document}